\documentclass[english]{article}
\usepackage[T1]{fontenc}
\usepackage[latin9]{inputenc}
\usepackage{float}
\usepackage{amsmath}
\usepackage{amsthm}
\usepackage{amssymb}
\usepackage{stackrel}

\makeatletter

\floatstyle{ruled}
\newfloat{algorithm}{tbp}{loa}
\providecommand{\algorithmname}{Algorithm}
\floatname{algorithm}{\protect\algorithmname}

\numberwithin{equation}{section}
\numberwithin{figure}{section}

\@ifundefined{date}{}{\date{}}
\makeatother

\usepackage{babel}
\begin{document}

\title{Cryptanalysis of Xinyu et al.'s NTRU-Lattice Based Key Exchange Protocol}

\author{{\small{}Maheswara Rao Valluri}\\
{\small{}School of Mathematical \& Computing Sciences }\\
{\small{}Fiji National University}\\
{\small{}P.O.Box No:7222, Derrick Campus, Suva,Fiji}}
\maketitle
\begin{abstract}
Xinyu et al. proposed a public key exchange protocol, which is based
on the NTRU-lattice based cryptography. In this paper, we show how
Xinyu et al.'s NTRU-KE: A lattice based key exchange protocol can
be broken, under the assumption that a man-in-the middle attack is
used for extracting private keys of users who participate in the key
exchange protocol. 
\end{abstract}

\section{Introduction}

Key exchange protocol is one of the public key cryptographic primitives
that provides a platform to negotiate keys among a group of parties.
In a key exchange protocol, private keys can be exchanged among the
group of parties over public insecure communication networks and agree
upon a common session key, which can be used for later secure communication
among them. A security goal of the key exchange protocol is that private
keys are to be shared among the group of parties without compromising
their secrecy. Key exchange protocol works as one of the basic building
blocks for constructing other high-level secure protocols and is used
to provide perfect forward secrecy in Transport Layer Security's ephemeral
modes.

A key exchange protocol was introduced by Diffie-Hellman during 1976
in their seminal paper {[}8{]}. The Diffie-Hellman (or Elliptic Curve
Diffie-Hellman) key exchange protocol security relies on the (or Elliptic
Curve) discrete logarithmic problem over a finite field {[}4,6,8{]}.
The Diffie-Hellman (or Elliptic Curve Diffie-Hellmann) key exchange
protocol is vulnerable to a quantum computer attack as a result of
the Shor's algorithm {[}5{]}. Recent proposals {[}1,2,7{]} for key
exchange protocol are alternative candidates against quantum attack.
In 2013, Xinyu et al. proposed a public key exchange protocol {[}7{]}
which is no longer dependent on number-theoretic cryptographic hard
assumption problems. Their proposal is based on the post-quantum NTRU-lattice
based cryptosystem {[}3{]}. The NTRU lattice based encryption is a
fast feasibly secured scheme introduced in 1996 by Hoffstein et al.
{[}3{]} and has been commercially standardized. The security of the
NTRU cryptosystem {[}3{]} relies on the ``shortest vector problem''
and the ``closest vector problem''. Xinyu et al. proposed an NTRU
- key exchange protocol {[}7{]} by adapting the NTRU lattice based
encryption scheme {[}3{]}.

A man-in the middle attack is a cyber attack where an eavesdropper
intercepts a communication between two participated parties and injects
his own message. This attack can succeed only when the eavesdropper,
Eve can impersonate each endpoint to their satisfaction as expected
from the legitimate other end. This paper describes how Xinyu et al.'s
NTRU-Key exchange protocol is not safe against man -in-the middle
attack; Eve is able to recover both parties' private keys.

\section{Preliminaries}

\subsection{Notations and Mathematical Background}

Throughout this paper, we denote $\mathbb{Z}$ the integer ring and
$\mathbb{Z}_{q}$ is the ring $\mathbb{Z}/q\mathbb{Z}$. A truncated
polynomial ring $R_{q}=\mathbb{Z}_{q}[x]/(x^{N}-1)$ consists of polynomials
with degree less than $N$ and co-efficients in $\mathbb{Z}_{q}$.
An element $f\in R_{q}$ can be written as a polynomial,

$f=\stackrel[I=0]{N-1}{\sum}f_{i}x^{i}=[f_{0},f_{1},.....,f_{N-1}]$.

Two polynomials $f$ and $g\in R_{q}$ are multiplied by the ordinary
convolution,

$(f*g)_{k}=\stackrel[i+j\equiv k(mod\,N)]{}{\sum}(f_{i}.g_{j}),k=0,1,...,N-1,$
which is commutative and associative. The convolution product is presented
by {*} to distinguish it from the multiplication in $\mathbb{Z}_{q}$.
We define a center $l_{2}-$norm of an element $f\in R_{q}$ by $\Vert f\Vert_{2}=(\stackrel[i=0]{N-1}{\sum}f_{i}-\bar{f})^{1/2}$,
where $\bar{f}=\frac{1}{N}\stackrel[i=0]{N-1}{\sum}f_{i}$.

\section{Xinyu et al.'s NTRU-Key exchange protocol}

Xinyu et al.'s NTRU - Key exchange protocol {[}7{]} public parameters
(N,p,q) are chosen such that N- is a prime, p and q are co-prime,
$gcd(p,q)=1$, and q is larger than p. The selection process of parameters
is the same as the NTRU-Encryption scheme {[}3{]}. Readers are encouraged
to refer to {[}7{]} for further details. We now present briefly Xinyu
et al.'s NTRU - Key exchange protocol in the following algorithm 1.

\begin{algorithm}
\caption{Xinyu et al.'s NTRU-Key Exchange Protocol}

\textbf{Alice} ~~~~~~~~~~~~~~~~~~~~~~~~~~~~~~~~~~~~~
~~~~~~~~~~~~~~~~~~\textbf{\small{}~Bob}{\small \par}

\textbf{\textit{\small{}Step 1:}}{\small \par}

$f_{A}\xleftarrow{\$}\mathfrak{L}_{f}$,$g_{A}\xleftarrow{\$}\mathfrak{L}_{g}$
~~~~~~~~~~~~~~~~~~~~~~~~~~~~~~~~~~~~~~\textbf{\textit{\small{}~Step
2:}}{\small \par}

$h_{A}=f_{A}^{-1}*g_{A}(mod\,q)$ ~~~ ~~ ~~~~~~~~~~$\underrightarrow{h_{A}}$~~~~~~~~~~~~~~~$f_{B}\xleftarrow{\$}\mathfrak{L}_{f}$,$g_{B}\xleftarrow{\$}\mathfrak{L}_{g}$
,$r_{B}\xleftarrow{\$}\mathfrak{L}_{r}$

\textbf{\textit{\small{}Step3:}}~~~~~~~~~~~~~~~~~~~~~~~~~~~
~~~~~~~~~$\underleftarrow{h_{B,}e_{B}}$~~~~~~~~~~$e_{B}=pr_{B}*h_{A}+f_{B}(mod\,q)$

$r_{A}\xleftarrow{\$}\mathfrak{L}_{r}$ 

$e_{B}=pr_{B}*h_{A}+f_{B}(mod\,q)$ ~~~~~~~~~~$\underrightarrow{e_{B}}$~~~~~~~~~~~~~~~~\textbf{\textit{\small{}Step
4:}}{\small \par}

$i_{A}=f_{A}*e_{B}(mod\,q)$ ~~~~~~~~~~~~~~~~~~~~~~~~~~~~~~~~~~~~~$i_{B}=f_{B}*e_{A}(mod\,q)$

$K_{A}=i_{A}(mod\,p)=f_{A}*f_{B}(mod\,p)$~~~~~~~~~~~~~~~~~~~~$K_{B}=i_{B}(mod\,p)=f_{B}*f_{A}(mod\,p)$
\end{algorithm}

\section{Man-in-the middle attack on Xinyu et al.'s NTRU-Key exchange Protocol}

In this section, we show weakness of Xinyu et al.'s NTRU-Key exchange
protocol against man-in-the middle attacks. To show this Xinyu et.al's
protocol is insecure against a man-in the middle attack, let us suppose
an adversary, named Eve listens to communication between both parties,
Alice and Bob who believe they are communicating with each other. 

A man-in-the middle attack is described as follows: 

\textbf{\textit{-Step 1: }}An adversary, Eve intercepts both the public
keys $h_{A}$ and $h_{B}$ sent by Alice and Bob, respectively and
computes two of her own public keys, $h^{'}=f^{'-1}*g^{'}(mod\,q)$
and $h^{''}=f^{''-1}*g^{''}(mod\,q)$ such that there exist the inverse
of $f^{'}$ and $f^{''}$ in $R_{p}$ and $R_{q}$. Then, she sends
$h^{'}$ to Alice and $h^{''}$ to Bob. 

\textbf{\textit{-Step 2: }}After receiving $h^{'}$, Alice picks $r_{A}\xleftarrow{\$}\mathfrak{L}_{r}$
and computes $e^{'}=pr_{A}*h^{'}+f_{A}(mod\,q)$ and sends $e^{'}$
to Bob, but Eve intercepts it. Similarly, after receiving $h^{''}$,
Bob picks $r_{B}\xleftarrow{\$}\mathfrak{L}_{r}$ and computes $e^{''}=pr_{B}*h^{''}+f_{B}(mod\,q)$
and sends $e_{B}$ to Alice, but Eve intercepts it.

\textbf{\textit{Step 3:}} After intercepting $e^{'}$ and $e^{''}$
sent by Alice and Bob, respectively, Eve computes $w_{A}=e^{'}*f^{'}(mod\,q)$,
$K_{A}^{'}=w_{A}*f^{'-1}(mod\,p)=f_{A}(mod\,p)$ and similarly, $w_{B}=e^{''}*f^{''}(mod\,q)$,
$K_{B}^{'}=w_{B}*f^{''-1}(mod\,p)=f_{B}(mod\,p)$ to recover Alice's
private key $f_{A}$ and Bob's private key$f_{B}$, respectively.

\section{conclusion}

In this paper, we have described how Xinyu et al.'s NTRU-KE protocol
is insecure against a man-in-the middle attack. The process of this
attack has been explained-how an attacker can recover private keys
of both parties. Note that Xinyu et al.'s NTRU-Key exchange protocol
is trivially insecure against a man-in the middle attack for unauthenticated
public keys used in this protocol.

\end{document}